\documentclass[twocolumn,aps,a4paper,superscriptaddress,showpacs,showkeys,prl,reprint]{revtex4-1}
\usepackage{graphicx}
\usepackage{subfigure}
\usepackage{subfigure}
\usepackage{latexsym}   
\usepackage{amsmath,amssymb,amsthm}

\usepackage{color}

\begin{document}




\title{Importance of non-affine viscoelastic response in disordered fibre networks}

\author{\firstname{L.} G. \surname{Rizzi}} %
\affiliation{Departamento de F\'isica, Universidade Federal de Vi\c{c}osa, 36570-900, Vi\c{c}osa, MG, Brazil.}
\affiliation{School of Chemistry, University of Leeds, LS2 9JT, Leeds, UK.}
\author{\firstname{S.} \surname{Auer}} %
\affiliation{School of Chemistry, University of Leeds, LS2 9JT, Leeds, UK.}
\author{\firstname{D.} A. \surname{Head}} %
\affiliation{School of Computing, University of Leeds, LS2 9JT, Leeds, UK.}

\date{\today}


\begin{abstract}
\noindent
	Disordered fibre networks are ubiquitous in nature and have a wide range of industrial applications as novel biomaterials. 
	Predicting their viscoelastic response is straightforward for affine deformations that are uniform over all length scales, but when affinity fails, as has been observed experimentally, modelling becomes challenging. 
	Here we introduce a numerical methodology to predict the steady-state viscoelastic spectra and degree of affinity for disordered fibre networks driven at arbitrary frequencies. Applying this method to a peptide gel model reveals a monotonic increase of the shear modulus as the soft, non-affine normal modes are successively suppressed as the driving frequency increases. In addition to being dominated by fibril bending, these low frequency network modes are also shown to be delocalised. 
	The presented methodology provides insights into the importance of non-affinity in the viscoelastic response of peptide gels, and is easily extendible to all types of fibre networks.
\end{abstract}



\maketitle

\noindent


	Fibrous assemblies represent an important class of materials with many industrial applications including scaffolds for tissue engineering~\cite{BurdickBook} and enamel remineralization~\cite{Brunton2013}, nonwoven fabrics for medical textiles and industrial filters~\cite{RussellBook}, carbon nanotube composites~\cite{Hall2008}, paper and felt~\cite{Alava2006}.
	Nature employs protein fibre networks in the multi-functional cellular cytoskeleton~\cite{AlbertsBook,BrayBook}.
	The mechanical stiffness of fibre networks is often central to their function, and although static properties come under most scrutiny, they often exist in dynamic environments subject to temporally-varying mechanical loads, including the cytoskeleton of motile cells~\cite{BrayBook}, and scaffolds for tendon and ligament regeneration, where habitual loading propagating through the network influences the viability of embedded stem cells~\cite{Burdick2009,Raif2005,Appelman2011}. Understanding the dynamical network response is essential to design novel materials with properties suited for such situations.

\begin{figure}[!t]
\centering
\includegraphics[width=0.43\textwidth]{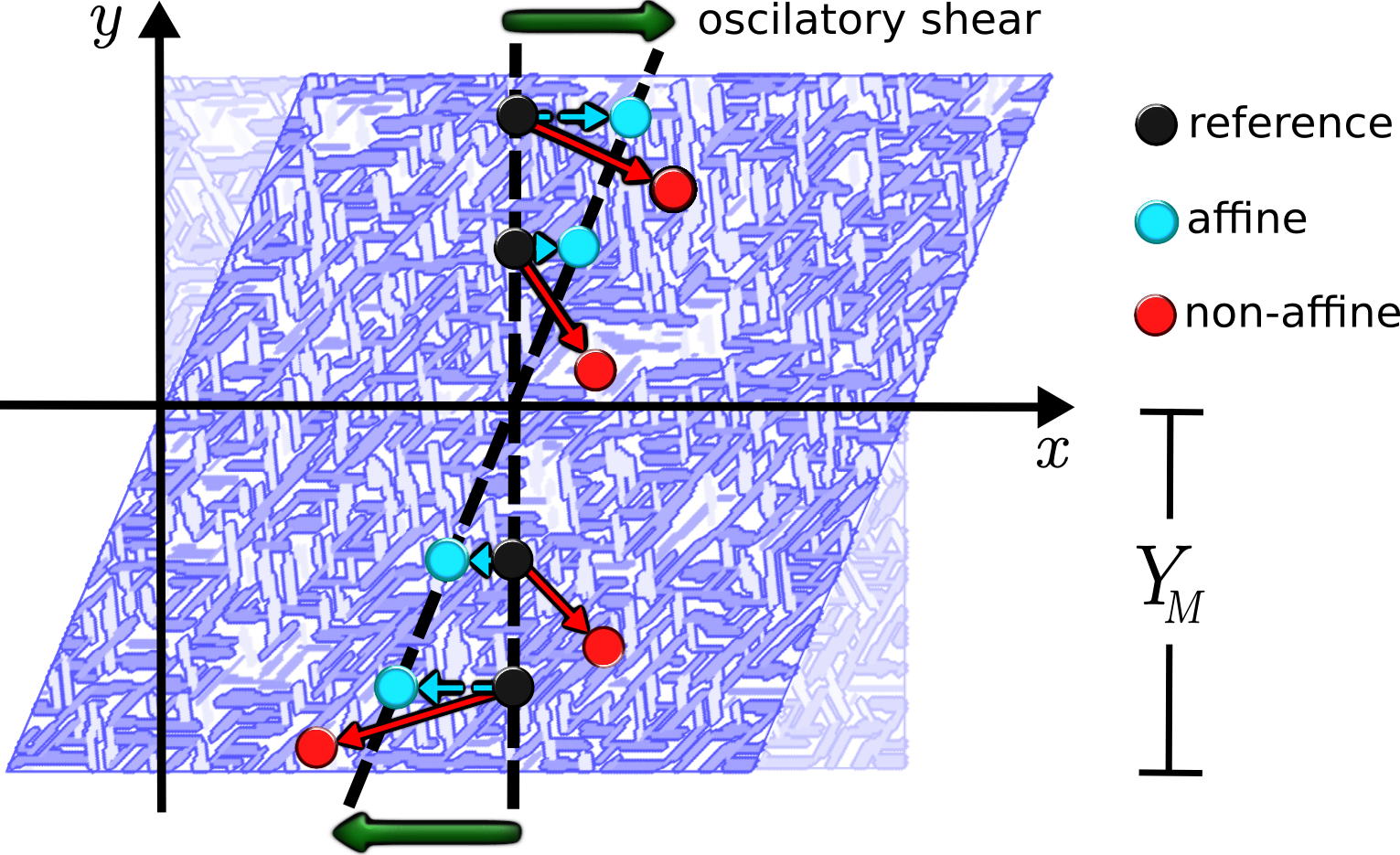}
\caption{Schematic representation of affine (light discs and arrows) and non-affine (dark discs and arrows) deformations on a fibre network under shear. On the background is a fibre network configuration extracted from our simulations. Fibres are formed from the self-assemble of anisotropically interacting peptide monomers~\cite{rizzi2015jchemphysB,rizzi2015prl}.}
\label{nonaffine_def}
\end{figure}

	A key modelling challenge is to determine the degree to which the deformation is affine~\cite{wen2012softmat}, {\em i.e.}~uniform over all relevant length scales; see Fig.~\ref{nonaffine_def}.
	If affinity holds, extrapolating the macroscopic response from a putative microstructure is straightforward, and a range of thermal and athermal affine models for fibre networks have been developed~\cite{Broedersz2014,Pritchard2010}.
	When affinity fails, however, as experimentally observed over broad parameter ranges~\cite{Gardel2004,Liu2007,Piechocka2011,Atakhorrami2014}, it is necessary to determine the microscopic deformation field, which typically requires numerical solution for explicit network realisations.
	This has thus far been limited to the elastic plateau amenable to energy minimization algorithms~\cite{Head2003,Wilhelm2003,Buxton2007,Astrom2008}, or computationally--intensive particle methods that only access short times~\cite{Kim2009,Huisman2010}.
	Without a more general understanding of fibre networks dynamics, we lack the capability to predict potentially large changes in viscoelastic properties over experimentally relevant time scales.

Here we present a methodology which allows the numerical calculation of the viscoelastic spectra for any type of disordered fibre network driven at arbitrary oscillation frequencies. The method is based on normal modes which ensures linear response, and since no thermal effects or crosslink dynamics are included by construction, all measured variation in affinity and viscoelasticity can be ascribed with certainty to network properties. We demonstrate the efficacy of this method by applying it to a model of peptide gels, and reveal a rich interplay between viscoelasticity, affinity, and mode localisation that derives from the successive suppression of network modes as the driving frequency increases.

	Our considerations apply to crosslinked networks of slender elastic fibres immersed in a Newtonian fluid with viscosity $\nu$.
	To simplify the network-fluid interaction, all fibre mass is regarded as being concentrated on network nodes in the form of a spherical bead with radius~$a$ and corresponding Stoke's drag coefficient $6\pi a\nu$.
	Hydrodynamic interactions between beads are neglected.
	Taking the overdamped regime relevant to the intended applications, the force balance equation in terms of the node/bead displacement  $\vec{u}$ is
\begin{equation}
6 \pi a \nu \partial_t \vec{u} + H \vec{u} = \vec{f} \cos(\omega t) ~~,
\label{force:eq}
\end{equation}
where $H$ is the dynamical (Hessian) matrix with components $H_{ij}\equiv\partial_{i}\partial_{j}E_{\text{elastic}}$ in terms of the total elastic energy $E_{\rm elastic}(\{\vec{u}\})$ 
of a given configuration, and $\vec{f}$ is the vector amplitude of the force applied to this node.
	The left hand side of (\ref{force:eq}) couples fluid friction to internal forces generated by network elasticity, and these are balanced with the external force on the right hand side, here assumed to be oscillatory.
	A stress-controlled shear protocol is assumed where the force is applied only to boundary nodes, so that $\vec{f}=0$ for the internal nodes, $\vec{f}=+\vec{f}_0$ on upper boundary nodes, and $\vec{f}=-\vec{f}_0$ on lower boundary nodes, where all $\vec{f}_0$ on each surface sum to give the required stress.
	All node displacements are indexed into a single vector $\vec{U}$, which could be ordered {\it e.g.}
($u_{1,x}, u_{1,y}, u_{2,x} , u_{2,y} , \dots$) for two dimensional (2D) networks.
	All node displacements can then be written in terms of the eigenvectors $\vec{h}_{\alpha}$ of the Hessian $H$ as
\begin{equation}
\vec{U} = \sum_{\alpha} \bar{u}_{\alpha} \vec{h}_{\alpha} ~~,
\label{uexpansion}
\end{equation}
where the sum is over all modes~$\alpha$.

	By substituting the expansion above into (\ref{force:eq}) we obtain exact expressions for the in-phase $\bar{u}^{\prime}_{\alpha}$ and out-of-phase $\bar{u}^{\prime \prime}_{\alpha}$ components of the coefficients $\bar{u}_{\alpha}$ in steady state, 
\begin{equation}
\bar{u}^{\prime}_{\alpha}(t) = \frac{1}{1 + (\omega \tau_{\alpha})^2} \frac{\bar{f}_{\alpha}}{\lambda_{\alpha}} \cos(\omega t)
\label{uprime}
\end{equation}
and
\begin{equation}
\bar{u}^{\prime \prime}_{\alpha}(t) = \frac{\omega \tau_{\alpha}}{1 + (\omega \tau_{\alpha})^2} \frac{\bar{f}_{\alpha}}{\lambda_{\alpha}} \sin(\omega t) ~~,
\label{udoubleprime}
\end{equation}
where $\bar{f}_{\alpha}$ are the coefficients of the expansion $\sum_{\alpha} \bar{f}_{\alpha} \vec{h}_{i,\alpha}$ for the external force on all nodes, and $\lambda_{\alpha}$ is the eigenvalue of mode~$\alpha$.
	The eigenvalues $\lambda_{\alpha}$ are usually related to frequencies, but because we consider the overdamped limit they are instead related to relaxation times $\tau^{\text{sim}}_{\alpha}=6 \pi a \nu/\lambda_{\alpha}$.
	Note that floppy modes correspond to null eigenvalues and undefined relaxation times.
	We identify these using singular value decomposition~\cite{NumericalRecipes}, and assign to each the coefficients $\bar{u}^{\prime}_{\alpha}=0$ and $\bar{u}^{\prime \prime}_{\alpha}=\omega^{-1} (\bar{f}_{\alpha}/6 \pi a \nu ) \sin(\omega t)$ corresponding to $H\vec{u}=\vec{0}$ in~(\ref{force:eq}).
	By considering the amplitudes in (\ref{uprime}) and (\ref{udoubleprime}), one can use (\ref{uexpansion}) to relate the displacements $\vec{u}_{i}$ to the local strain in the $i$-th bead as $\gamma_{i}=u_{i,x} / (u_{i,y}-Y_M)$, where $Y_M$ is the middle height line of the system (see Fig.~\ref{nonaffine_def}).
	In order to avoid numerical instabilities due to those beads near the middle line ({\it i.e. $u_{i,y} \approx Y_M$}), we take the mean value averaged only over beads placed at the upper and bottom boundaries.
	Finally, the in-phase ($\gamma^{\prime}$) and out-of-phase ($\gamma^{\prime \prime}$) strains are used to compute the shear moduli of the fibre network, {\it i.e.}~both the storage modulus $G^{\prime}(\omega)=\left\langle f_{0} / \gamma^{\prime} \right\rangle$ and the loss modulus $G^{\prime \prime}(\omega)=\left\langle f_{0} / \gamma^{\prime \prime}  \right\rangle$.

	In practice, the numerical determination of the viscoelastic spectra of a disordered fibre network requires (i) the construction of the Hessian matrix $H$ for an explicit network realisation and a chosen model for single-fibril elasticity, and the determination of its eigenvectors $\vec{h}_{\alpha}$ and eigenvalues $\lambda_{\alpha}$, (ii) the determination of the coefficients $\bar{f}_{\alpha}$ in the expansion of the external force on the network nodes in terms of the eigenvectors, (iii) knowledge of $\tau_{\alpha}$, $\lambda_{\alpha}$ and $\bar{f}_{\alpha}$ allows determination of the in-phase and out-phase response $\bar{u}^{\prime}_{\alpha}$ and $\bar{u}^{\prime \prime}_{\alpha}$ from (\ref{uprime}) and (\ref{udoubleprime}), which in turn allows determination of the actual displacement $\vec{U}$ from (\ref{uexpansion}) as a function of the frequency~$\omega$, (iv) from $\vec{U}$ it is straightforward to determine the local strains $\gamma^{\prime}$, $\gamma^{\prime \prime}$ and the shear moduli  $G^{\prime}$, $G^{\prime \prime}$ of the fibre network from the above formulae.


	Our test system is a recently developed 2D model for peptide gels, where peptide monomers are explicitly considered in the formation of the fibre network~\cite{rizzi2015prl}, which generalises a lattice-based elastic network model~\cite{broedersz2011np} to permit variations in fibre thickness.
	The interactions between peptide monomers are characterized by their anisotropy ratio $\xi = \psi / \psi_h > 1$, where $\psi$ and $\psi_h$ are the strengths of strong directional hydrogen bonds and weak isotropic hydrophobicity-mediated bonds~\cite{rizzi2015jchemphysB}, respectively.
	The anisotropy in the interactions between peptide monomers enables their assembly into crosslinked networks that exhibit a universal time-dependent behaviour in their microstructural geometry ({\it i.e.}~fibre thickness, fibre length, crosslink separation).
	Furthermore, the same time-scaling function was found to collapse the plateau value of the corresponding shear modulus and crosslink connectivities~\cite{rizzi2015prl} .


	Unless otherwise stated, all results presented below are for networks generated from monomers with anisotropy $\xi=10$ and a coverage (mean lattice occupation) $\theta=0.525$, obtained at two different simulation times $t$ measured in Monte Carlo steps (MCS).
	All measurements correspond to averages over 25 independent simulations.
	Results are reported in experimental units assuming a Young's modulus for the fibrous material to be $E^{f}=10^{9}$~Pa, all beads having the same radius $a=10$~nm, and the fluid viscosity $\nu=0.001$~Pa~s is that of water at 20$^{\text{o}}$C.
	Simulation relaxation times and frequencies are converted to experimental units as per $\tau = \tau^{\text{sim}}/E^f$ and $\omega = \omega^{\text{sim}}E^f$, with units of s and s$^{-1}$ respectively.
	In addition, the viscoelastic spectra $G^{\prime}(\omega)$ and $G^{\prime\prime}(\omega)$ have been normalised to the frequency-independent affine shear modulus $G_{\rm aff}$ corresponding to the storage modulus at zero frequency.



\begin{figure}[!t]
\centering
\includegraphics[width=0.49\textwidth]{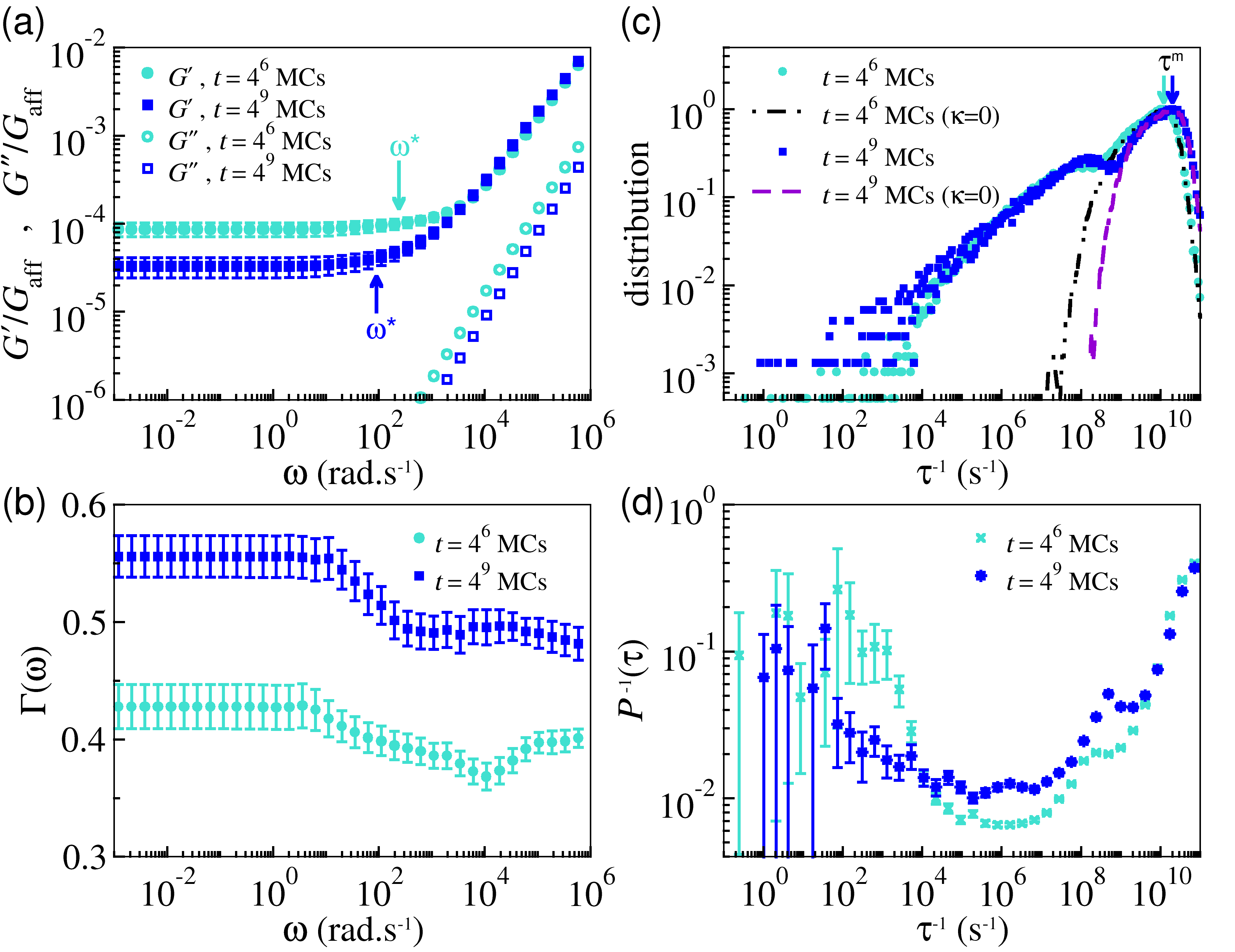}
\caption{Results obtained for a fibre network with $34\,408$ peptides ($\theta=0.525$) and anisotropic ratio $\xi=10$ at two simulation times $t=4^6$~MCs and $t=4^9$~MCs.
(a) normalized storage modulus $G'(\omega)/G_{\rm aff}$ (filled symbols) and loss modulus $G''(\omega)/G_{\rm aff}$ (open symbols).
(b) non-affinity parameter $\Gamma(\omega)$.
(c) distribution of relaxation times $\tau$, where $\kappa=0$ denotes distributions obtained neglecting the bending terms of the elastic energy.
(d) Inverse of the participation ratio $P^{-1}(\tau)$. Error bars are computed as the standard deviation from 25 independent simulations.}
\label{shear_main_results}
\end{figure}

	Figure~\ref{shear_main_results}(a) demonstrates that the storage modulus $G'$ presents a plateau regime for low frequencies, and then smoothly increases above some threshold frequency here denoted $\omega^{*}$.
	This behaviour can be rationalised in terms of the frequency cut-offs, {\em i.e.} the $1+(\omega\tau_{\alpha})^{2}$ factors in the denominators of (\ref{uprime}) and (\ref{udoubleprime}), leading to a reduction in the amplitude of mode $\alpha$ as $\omega$ increases beyond this mode's natural relaxation time~$\tau_{\alpha}$.
	Without this mode's contribution, the strain is reduced, so the system stiffens.
	At high frequencies, the increase of the storage modulus can be described by a power-law $G^{\prime} \sim \omega^{\delta}$ with $\delta$ in the range $0.5$ to $0.9$ for all values of $\xi$ and $t$ assayed.
	This range includes the value $\delta \sim 0.60$ measured for fibrillar networks using passive microrheology~\cite{corrigan2009langmuir}.
	An exponent of 0.5 due to crosslink unbinding dynamics has been observed in experiments~\cite{lieleg2007prl1} and confirmed theoretically~\cite{broedersz2010prl}, but as our model includes no such relaxation mechanism this cannot be the origin of our~$\delta$.
	Similarly the $3/4$ exponent for the wormlike chain model~\cite{gittes1998pre} requires thermal undulations that are not present in our athermal, elastic fibres.

	At the low frequencies, our networks deform in a highly non-affine manner as evident in the low values of $G^{\prime}/G_{\rm aff}$.
	This non-affine response is independently confirmed by simultaneously plotting the non-affinity parameter $\Gamma(\omega)= \langle u_{y}^2 /( u_{x}^2 + u_{y}^2 ) \rangle$, which is zero for affine deformations. 
	As seen in Fig.~\ref{shear_main_results}(b), $\Gamma$ increases with decreasing frequency.
 	Our 2D results can be compared to 3D experiments by scaling according to the affine predictions for each dimension, {\it i.e.}~$G_{3D}/G_{2D} = 8 / (15 l_c)$, with the inter-crosslink length $l_c \sim 10$~nm~\cite{rizzi2015prl}.
	This yields values for the storage modulus $G^{\prime}$ at the plateau regime equal to $(700 \pm 100)$~Pa for $\omega < w^* \approx 240$~rad.s$^{-1}$ and $(400 \pm 100)$~Pa for $\omega < w^* \approx 90$~rad.s$^{-1}$ at $t=4^6$~MCs and $t=4^9$~MCs, respectively.
	These values are comparable to measurements for peptide gels such as amyloid tapes~\cite{aggeli1997nature,greenfield2010langmuir,tang2011langmuir} and spider silk~\cite{rammensee2006apphysA,gong2010softmatt}.
	Fig.~\ref{shear_main_results}(a) also demonstrates our networks soften with age, which has also been observed for crosslinked actin~\cite{lieleg2011nature} and can be related here to the increase in non-affinity, itself due to the reduced network connectivity as shown elsewhere~\cite{rizzi2015prl}.


	In Fig.~\ref{shear_main_results}(c) we show the distribution of relaxation times $\tau$, which confirms that the broad range over which $G^{\prime}$ decreases is related to a broad range of $\tau$ following a bimodal distribution.
	Previous work at zero frequency identified the fast and slow relaxation peaks with fibre stretching and bending modes, respectively~\cite{Huisman2011}, and we can confirm this holds for finite frequency by setting the fibre bending modulus $\kappa$ to zero in $E_{\rm elastic}$, which removes the slow relaxation modes without significantly altering the fast ones as shown in the figure.
	In addition, the fast stretching modes move to shorter relaxation times as the simulation time $t$ increases, in contrast to the slow bending modes which remain fixed, lending insight into the mechanism underlying the observed softening with age.
	The slow bending modes are also delocalised, in contrast to the localised fast stretching modes, as shown in Fig.~\ref{shear_main_results}(d) where is displayed the inverse participation ratio $P^{-1}(\tau)=\sum_{\tau} |\vec{h}_{\tau} . \vec{h}_{\tau} |^2 / | \sum_{\tau'} \vec{h}_{\tau'} . \vec{h}_{\tau'} |^2$, which is high for delocalised and low for localised modes~\cite{Huisman2011,silbert2009pre}.
	This trend is consistent with intuitive assertions made in recent vimentin experiments~\cite{Head2014}.

\begin{figure}[!t]
\centering
\includegraphics[width=0.48\textwidth]{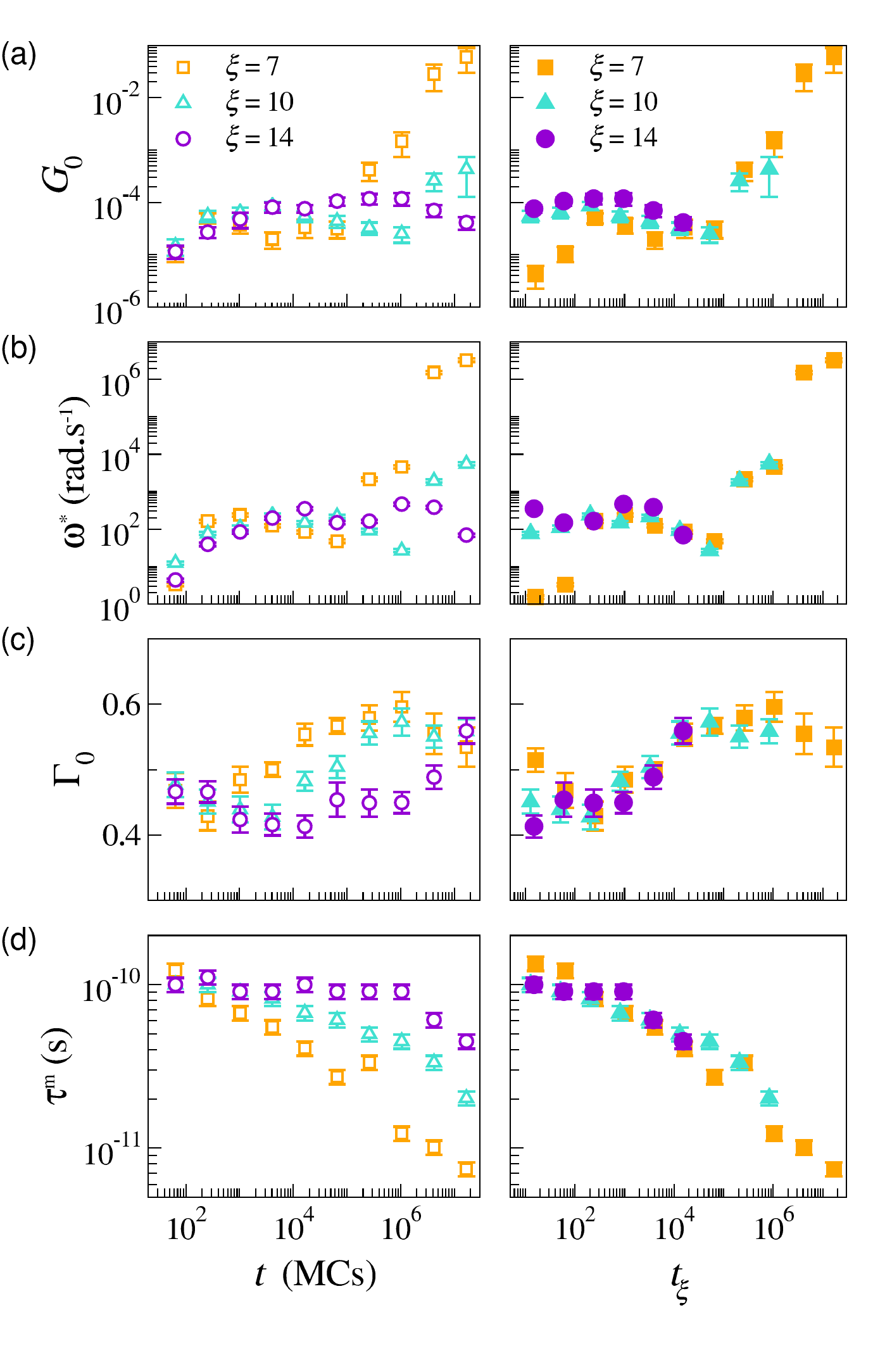}

\vspace{-0.2cm}

\caption{Time dependence of (a) the elastic modulus $G_{0}=G(0)$ scaled to the affine prediction, (b) the threshold frequency $\omega^*$, (c) the zero-frequency non-affinity $\Gamma_{0}$, and (d) the modal relaxation time $\tau^{m}$. The same quantities are plotted against the unscaled time $t$ (left panels, open symbols) and the rescales time $t_{\xi}$ (right panels, closed symbols).}

\label{scaling_rho525}
\end{figure}

\begin{figure}[!t]
\centering
\includegraphics[width=0.49\textwidth]{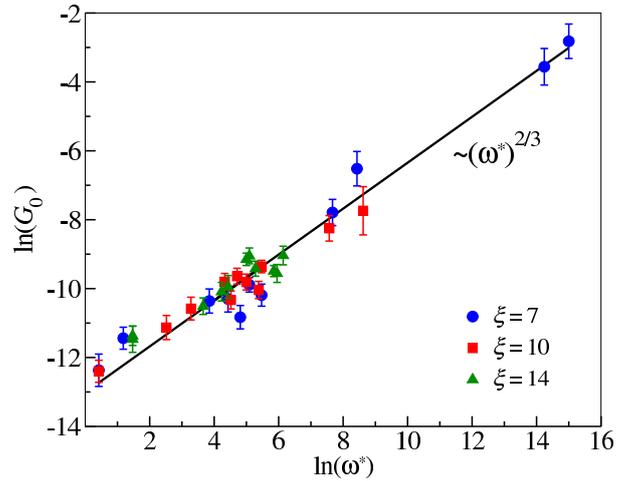}

\vspace{-0.2cm}

\caption{Power-law relation between the normalized plateau modulus $G_{0}$ and the threshold frequency $\omega^{*}$ for different anisotropy ratios $\xi$.}
\label{gomega_star}
\end{figure}

	The picture just described holds for other values of the anisotropy parameter $\xi$ and network formation time $t$ considered.
	Shown in Fig.~\ref{scaling_rho525} are the trends as $\xi$ and $t$ are varied for the zero-frequency elastic modulus $G_{0}\equiv G^{\prime}(\omega=0)/G_{\rm aff}$, the zero-frequency non-affinity $\Gamma_{0}\equiv\Gamma(\omega=0)$, the threshold frequency $\omega^{*}$ and the modal relaxation time $\tau^{m}$.
	In addition to the unscaled behaviour given as a function of simulation time $t$ (open symbols and left panels), we also plot the same quantities against the $\xi$-dependent rescaled time $t_{\xi} = t e^{-(\xi-\xi_0)}$ (filled symbols and right panels) which generates data collapse at zero frequency~\cite{rizzi2015prl}.
	As illustrated in Figs.~\ref{scaling_rho525}(a) and (b), $G_{0}$ and $\omega^{*}$ exhibit a similar non-monotonic behaviour, while the data for $\Gamma_{0}$ in Fig.~\ref{scaling_rho525}(c) demonstrates an increase in non-affinity with time.
	Figure~\ref{scaling_rho525}(d) confirms that the trend mentioned above, {\em i.e.} that $\tau^m$ shifts to shorter relaxation times with network age, is general.
	We also observe a power-law behaviour $G_{0} \sim (\omega^{*})^{2/3}$ which appears to be independent of $\xi$, as shown in Fig.~\ref{gomega_star}, but currently have no explanation for this apparently robust phenomenon.
	Finally, we can infer from the data collapse under the same rescaled time as~\cite{rizzi2015prl} that these dynamic quantities correlate to microstructural geometric quantities (fibre length and thicknesses, crosslink separation), suggesting the ultimate origin of the observed frequency dependence of our fibre networks is geometric.

	In summary, we have introduced an efficient numerical scheme to extract the linear finite-frequency viscoelastic response of fibre networks, and applied it to model peptide gels to observe a power-law increase of the storage modulus $G^{\prime}$ with frequency $\omega$.
	Our method precludes the possibility that this stiffening is related to dynamic crosslink unbinding~\cite{lieleg2007prl1,broedersz2010prl} or frequency-dependent single fibre response~\cite{gittes1998pre}, but instead demonstrates it is due to an underlying decrease in non-affinity as shown in Fig.~\ref{shear_main_results}.
	This prediction is in principle experimentally testable~\cite{Liu2007}.
	That the transition from affine to non-affine response is gradual is consistent with Brownian dynamics~\cite{Huisman2010} and elastic spring networks~\cite{yucht2011softmat}, although our results include fibre bending and are unambiguously steady state.
	The loss modulus $G^{\prime\prime}(\omega)$ never strongly deviated from the purely viscous response~$\nu\omega$, in contrast to the clearly sublinear variation observed in many fibrous materials~\cite{Roberts2012,rombouts2013softmat,lin2010,Broedersz2014}.
	This deviation may be due hydrodynamic interactions, which could be incorporated into this framework by including interaction terms {\em via} Oseen tensors~\cite{DoiEdwards} in (\ref{force:eq}) to give a dense matrix equation.
	Finally, we note that even though we have applied this methodology to peptide gels in 2D, we expect our method and core findings to be applicable to fibre networks in general, including in three dimensions.
	Our methodology also allows a way to approach the complex and largely unexplored problem of hydrodynamic interactions in fibre networks.


	We thank D. Mizuno for early discussions regarding the procedure detailed here.
	L.G.R. acknowledges support from the Brazilian agency CNPq (Grant N$^{\text{o}}$ 245412/2012-3). 
	D.A.H. acknowledges support from the Biomedical Health Research Centre, University of Leeds, UK.


%

\end{document}